\documentclass{iaus}
\usepackage{graphicx}

\title[Stellar Evolutionary Models: empirical constraints] 
{Stellar Evolutionary Models: challenges from observations of stellar systems}

\author[S. Cassisi]   
{Santi Cassisi$^1$}

\affiliation{$^1$INAF - Osservatorio Astronomico di Collurania, Via M. Maggini, s.n., 64100 Teramo, Italy.
\break email: cassisi@oa-teramo.inaf.it}

\pubyear{2007}
\volume{241}  
\pagerange{1--8}
\date{?? and in revised form ??}
\jname{Stellar Populations as Building Blocks of Galaxies}
\editors{A. Vazdekis, et al., eds.}
\begin{document}

\maketitle

\begin{abstract}

We briefly review some constraints\footnote{Owing to the limited number of pages of present review, only a sub-sample of the
topics discussed during the talk are briefly summarized. For the interested readers we are pleased to send them upon request
the complete presentation file.} for stellar models in various mass regimes and evolutionary stages as provided
by observational data from spectroscopy to multi-wavelenghts photometry. The accuracy of present generation of
stellar models can be significantly improved only through an extensive comparison between theory and observations. 

\end{abstract}

\firstsection 
\section{Introduction}

Stellar evolution models are pivotal ingredients in order to understand the evolutionary properties
of the various Stellar Populations present in both resolved and un-resolved stellar systems, so that they 
play a fundamental role in assessing the \lq{nature}\rq\ and the contribution of the different
blocks that contribute to build the galaxies.

In view of this relevance, it is quite important to establish the level of accuracy and
reliability of present generation of stellar models. This can be achieved only by comparing theoretical predictions
with empirical constraints.

During the second half of the last century, stellar evolution theory has allowed to 
properly understand the \lq{meaning}\rq\ of the various
branches observed in the Color Magnitude Diagram (CMD) of both galactic globular clusters (GGCs) and open clusters. 
This notwithstanding for a long time these theoretical predictions
were accounted for with an uncritical approach. They were used at face value without accounting
for theoretical uncertainties and their effect in deriving estimates about cluster age and distances.
More recently, this approach to the theoretical framework drastically changed and more
critical assessments were adopted.
The motivations at the base of this change have to be searched both in the will of providing reliable
estimates of the systematic uncertainties affecting this kind of comparison and in the relevant advances made in the
observational techniques and in the \lq{Physics}\rq\ applied to stellar models.

On the observational side, in recent years, the impressive improvements achieved for both photometric
and spectroscopic observations, has allowed to collect data of an unprecedent accuracy, which provide 
at the same time a stringent test and a challenge for the accuracy of the models.

On the theoretical side, even if significant improvements have been achieved
in the determination of the Equation of State (EOS), opacities, nuclear cross sections, neutrino emission rates
that are all fundamental physical inputs for solving the stellar structure equations, residual
uncertainties do exist still as it is clearly testified by the
not negligible differences still existing among evolutionary results 
provided by different theoretical groups. 
At the same time, models computed with this updated physics have
been extensively tested against the latest observations, and this has also contributed
to increase the awareness that it is no more possible to neglect physical processes as radiative
levitation, rotation, magnetic fields, considered secondary physical mechanisms until few years ago.

A careful discussion of the uncertainties affecting stellar models for low-mass stars was early addressed by
Chaboyer (1995), who investigated the reliability of  
theoretical predictions concerning H-burning structures presently evolving 
in GGCs and, in particular on the accuracy of age predictions. Such an investigation 
has been extended to later evolutionary phases by Cassisi et al. 
(1998, 1999), Castellani \& Degl'Innocenti (1999), and Gallart et al. (2005). A discussion of the drawbacks of stellar
models for low-mass stars and their impact on the most used age, distance and chemical composition
indicators has been also provided by Cassisi (2005 and references therein); while the issue of the main uncertainties affecting 
the evolutionary properties of intermediate-mass stars has been reviewed by Cassisi (2004).

\section{Stellar models: the \lq{building blocks}\rq }\label{sec:model}

From the point of view of people using stellar models, they provide: i) evolutionary lifetimes that can be compared with
suitable star counts; ii) bolometric luminosity and effective temperature that once converted in useful magnitudes and
colors in various photometric systems by  using color-$T_{eff}$ relations and Bolometric Correction scales, can
be compared with empirical data, and iii) predictions about the surface chemical abundances that can be tested 
against spectroscopical measurements.

However, any user before accounting for these theoretical predictions should ask himself this fundamental question: How much
accurate and reliable are the predictions coming out from stellar models?

It is clear that the reliability of a stellar model depends mostly on the accuracy of the adopted physical inputs 
as well as on the physical processes accounted for as: atomic diffusion, levitation, rotation. 

\begin{figure}
\centering
\resizebox{7.3cm}{!}{\includegraphics{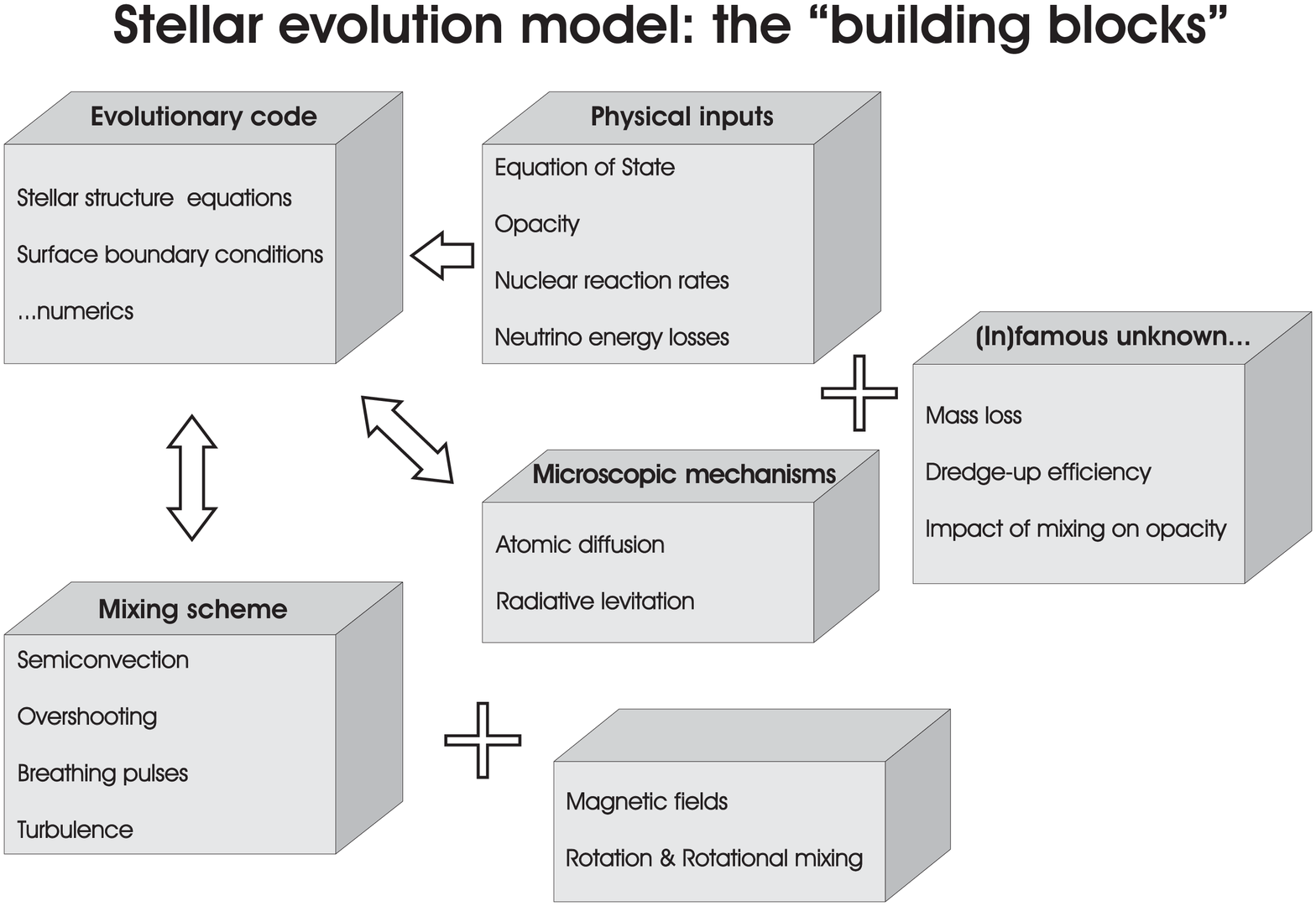} }
\caption[]{A qualitative view of the main \lq{blocks}\rq\ needed for \lq{building}\rq\ a stellar model.}
\end{figure}

In figure 1, we show a qualitative picture showing the most important \lq{building blocks}\rq\ that are required 
in order to construct a stellar model. 

The equations that describe the physical behavior of any stellar structure are well known since long time and a
detailed discussion on their physical meaning can be found in many books such as Cox \& Giuli (1968), Kippenhan \& Weigart
(1990) and Salaris \& Cassisi (2005). The solution of these equations is no more a problems - thanks also to the
advances in computing program and computer science -, as it has been shown by Weiss (2007, this conference) in the framework of
the \lq{Stellar Model Challenge}\rq: updated stellar evolution codes, 
once the physical scenario has been homogeneously fixed, provide results quite similar. 

However, in order to solve the stellar structure differential equations outer boundary conditions have to be provided: 
one can rely on some empirical relation for the thermal
stratification as that provided by Krishna-Swamy (1996) or a fully theoretical law as the so-called gray (or Eddington)
approximation, or alternatively one can adopt the predictions given by suitable model atmospheres. This issue for H-burning
stellar models has been recently reviewed by Vandenberg et al. (2007, and references therein) during this conference.

The meaning and role played by the various \lq{ingredients}\rq\ listed in fig. 1 has been extensively discussed in the
literature (see for instance Castellani 1999; Cassisi 2005). Some of them will be reviewed in the following in connection with the 
challenges provides by recent observations such as for diffusive processes or with recent advances in stellar physics as for the case of
conductive opacity.

Before closing this section, we wish to comment a bit on fig. 1. As already stated, this picture has only a qualitative
purpose. In the \lq{block}\rq\ {\it "(in)famous unknown..."} we put: mass loss efficiency, dredge-up efficiency and the impact of
mixing on opacity. When referring to the dredge-up efficiency we are considering the process occurring during the Asymptotic
Giant Branch (AGB).
In any case, for all the \lq{ingredients}\rq\ listed in this
block we are not yet able to predict their efficiency from fundamental principles and indeed we still rely on - quite
approximate - parametrization of the various processes.

In the block {\it "Additional physical processes"}, we include the presence of magnetic fields and the occurrence of rotation and
rotational-induced mixing. It is clear that in the implementation of both processes in an evolutionary code, due to our poor
knowledge of the physics at work would require a sizeable number of assumptions and free parameter. The reason for which we do
not include them in the {\it "(in)famous unknown..."} block is due to the evidence that we are always forced to account for 
mass loss and the $3^o$ dredge-up in order to explain the Horizontal Branch morphology (HB) and the evolution of AGB
stars; whereas we really need to account for magnetic field and rotation only to interpret some specific observational
features related, for instance, to the evolutionary properties of VLM stars and HB stars.

\section{Comparison theory - observations: the challenges}\label{sec:conf}

It is well known that depending of the stellar mass range, different physical processes affect the thermal and opacitive
properties of stellar structure. The accuracy of the adopted physical framework in different regimes can be tested by comparing
model predictions with various empirical data. 

\subsection{Very-Low-Mass Stars}

For many years, the computation of reliable models for Very-Low-Mass (VLM) stars has been severely challenged by the lack of robust
predictions about the thermal and opacitive properties as well as of suitable outer boundary conditions (Chabrier \& Baraffe 2000). 
As a consequence one was facing the tantalizing evidence that theoretical models were \lq{too blue}\rq\ to reproduce the observed
sequences of VLM both in clusters and in the field (Vandenberg et al. 1983).

In these last decade, on the theoretical side, the situation largely improved thanks to the recent availability of
appropriate EOS, radiative opacity, and outer boundary conditions
(Allard et al. 1997). From the observational point of view, thanks to the superb photometric capabilities of the Hubble
Space Telescope, a \lq{pletora}\rq\ of empirical data for such objects were collected (see King et al. 1998).
A plain evidence of the remarkable improvements achieved on this issue is represented by the nice fit to the faint MS of
the GGC NGC~6397 performed by VLM models by both Baraffe et al. (1998) and Cassisi et al. (2000).
Firm constraints for the theoretical framework are also provided by different types of empirical data as those given
by the Mass-Luminosity and Mass-Radius diagrams (see fig.~2).
The data showed in this plot reveal the existence of a very good agreement between theory and observations.

\begin{figure}
\centering
\resizebox{5.4cm}{!}{\includegraphics{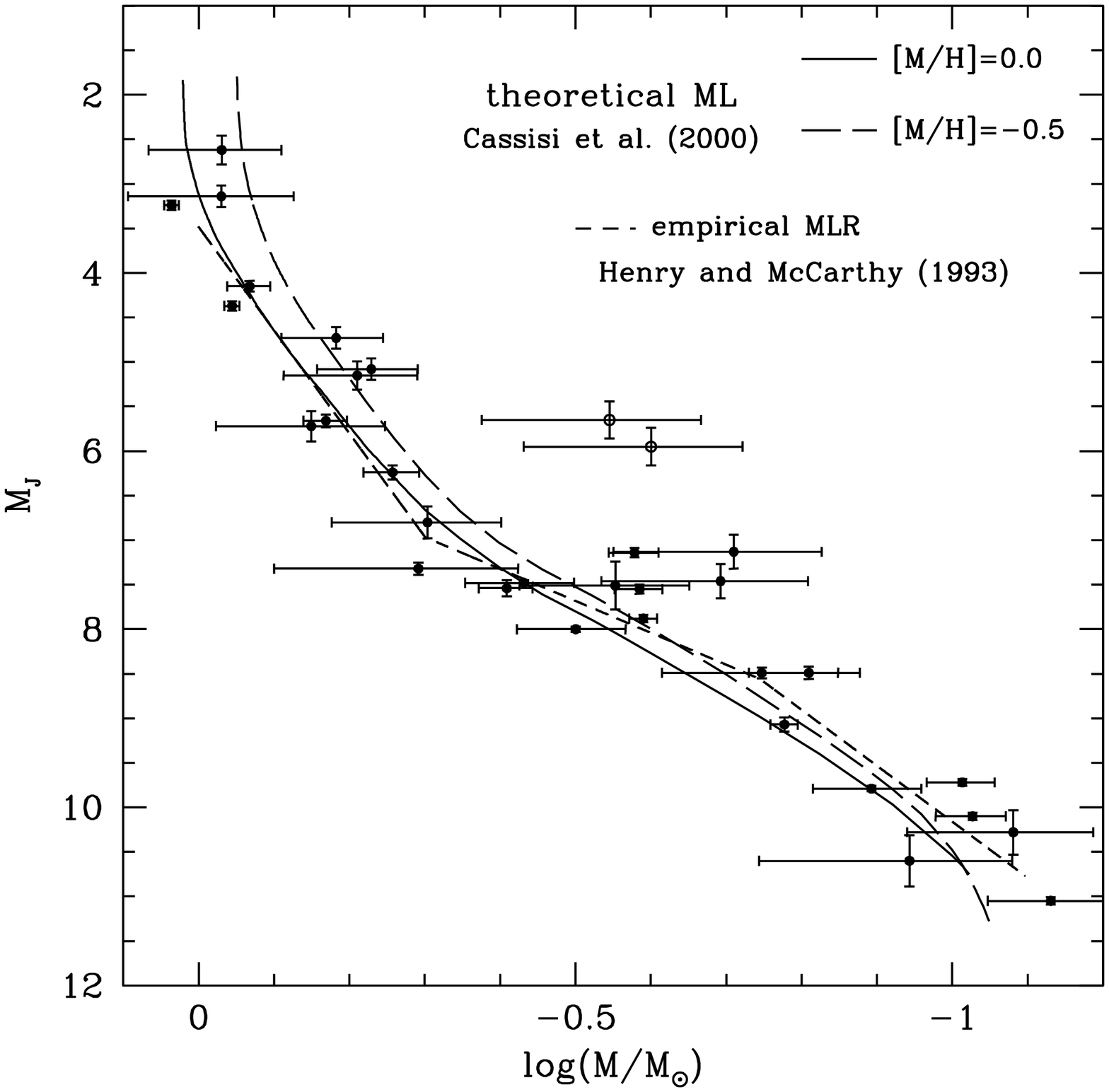} }
\resizebox{5.4cm}{!}{\includegraphics{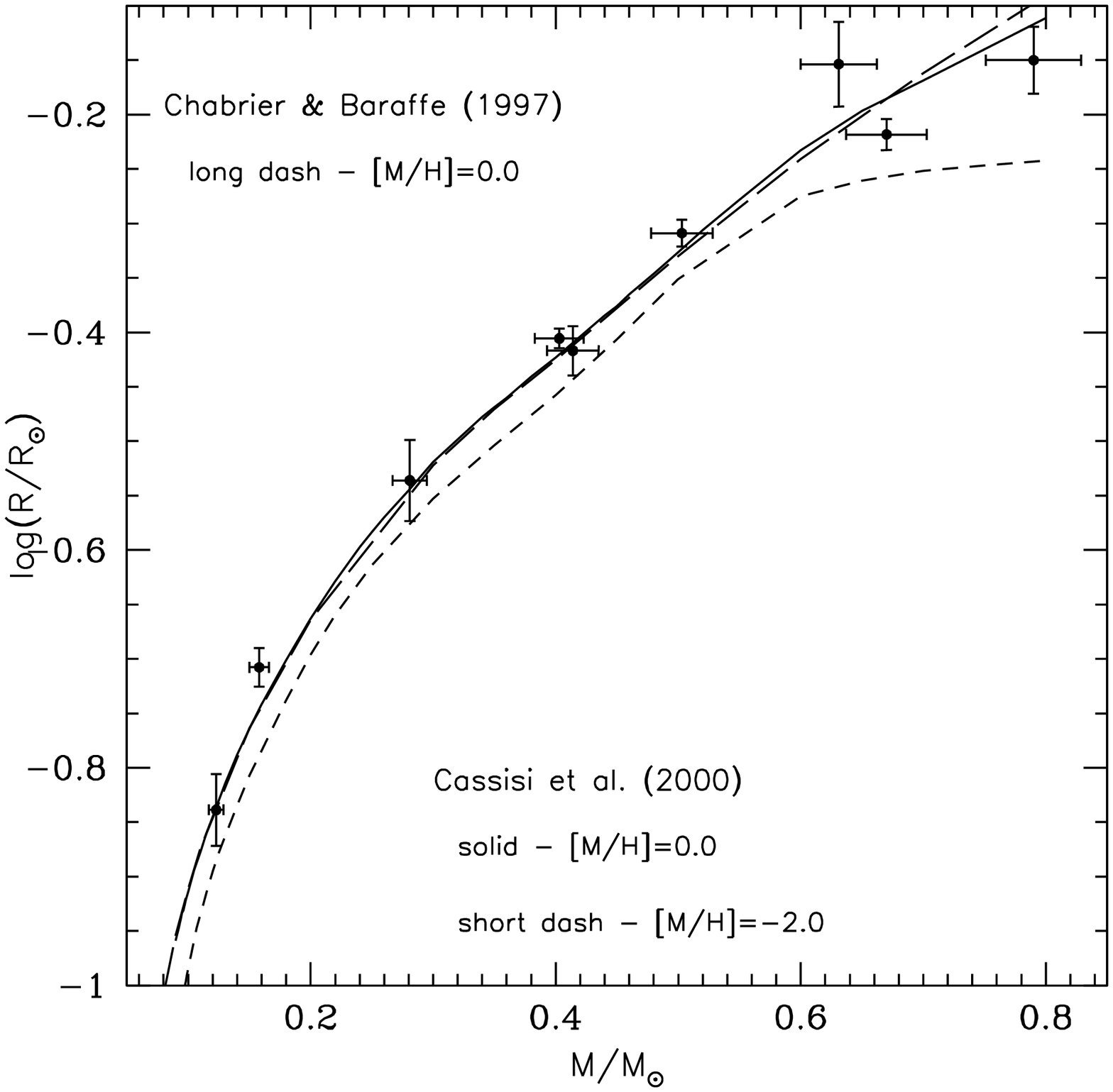} }
\caption[]{Left panel: comparison between VLM models and empirical data concerning the mass - luminosity relation.
Right panel: as left panel but for the mass-radius diagram (data from Segransan et al.~2003).}
\end{figure}

Therefore it seems that we can now be fully confident about the accuracy and reliability of VLM models; however
unfortunately this is not yet the case. The existence of shortcomings in these models appears
when taking into account empirical constraints represented by CMDs of intermediate- and metal-rich VLM stars. 
Left panel of fig.~3 shows the comparison
between VLM models and empirical optical data for the largest sample of field subdwarfs with known parallaxes: while models for
metal-poor composition finely reproduce the corresponding empirical sequence, the solar composition one clearly does not match
the data for $M_V>11$ mag. This evidence could be considered as a proof for a problem in the evolutionary models, however,
right panel of fig.~3 shows the comparison at longer wavelenghts between the same solar metallicity VLM models and empirical data for
field stars in the Bulge (Zoccali et al. 2000). It is worth noting that the same models that in the optical CMD do not fit
the data, in the Near-Infrared bands nicely reproduce the peculiar shape of the MS.

\begin{figure}
\centering
\resizebox{5.5cm}{!}{\includegraphics{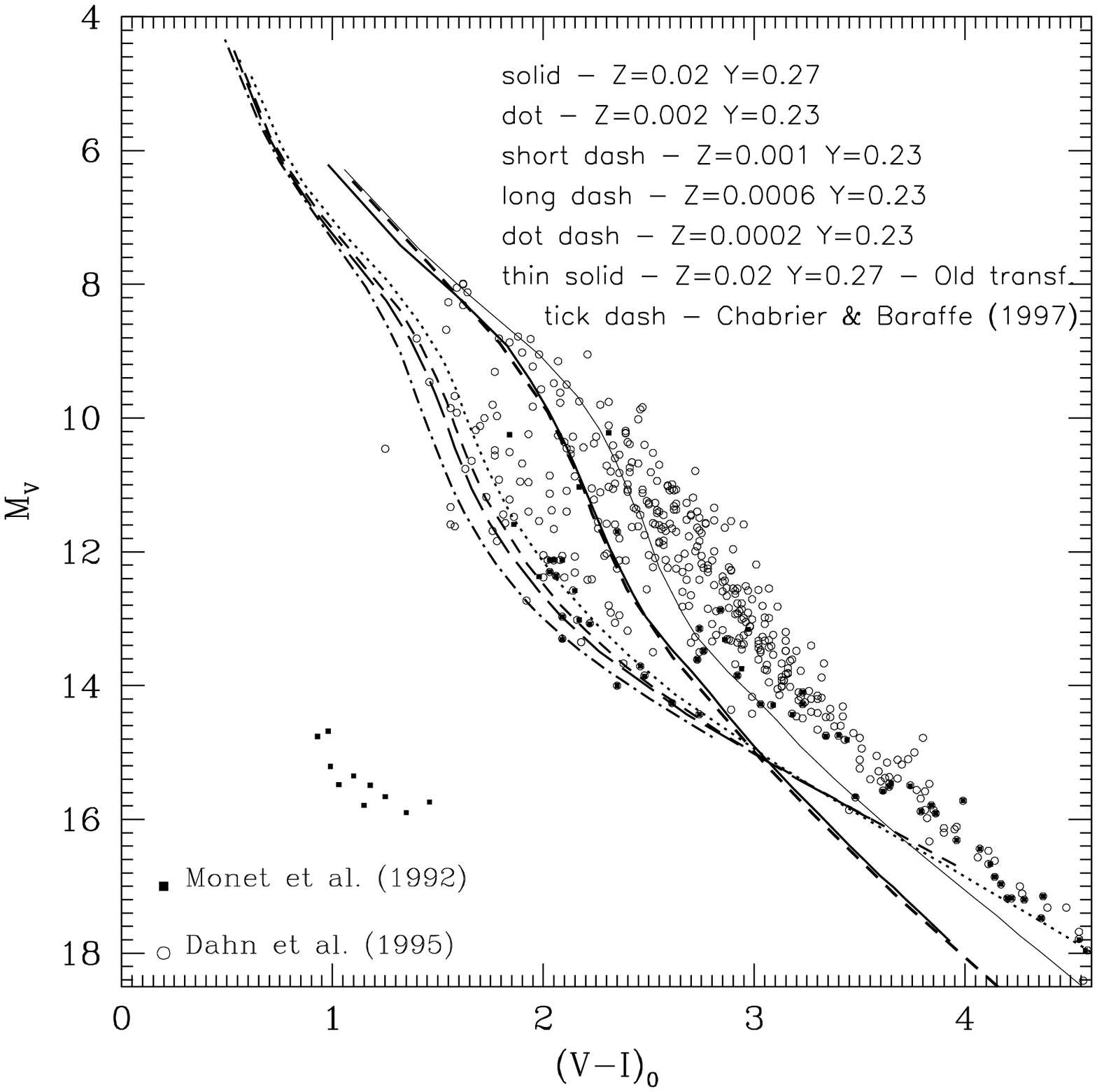} }
\resizebox{5.5cm}{!}{\includegraphics{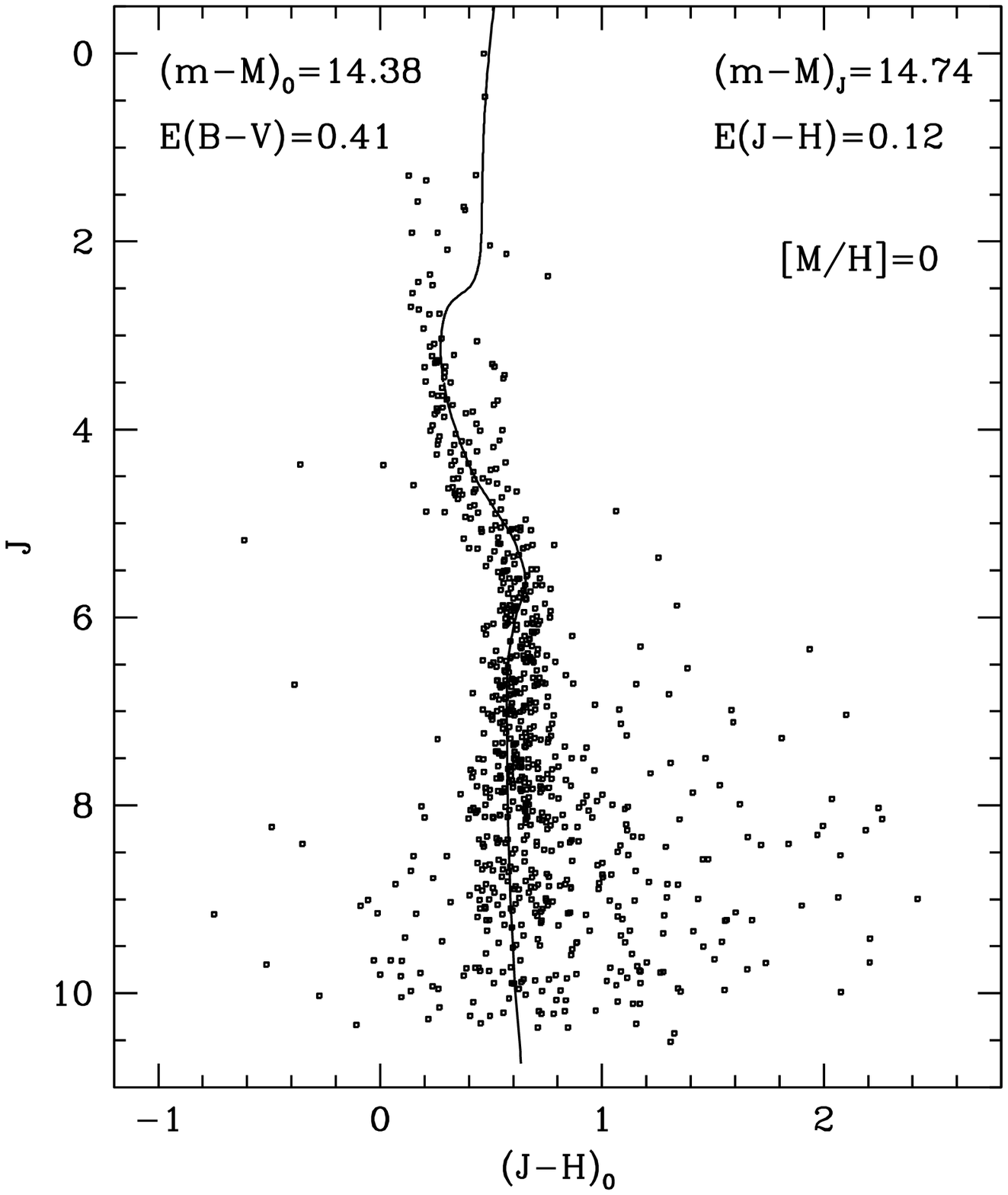} }
\caption[]{Left panel: comparison between VLM models (Cassisi et al. 2000) and the observed distribution of field dwarfs 
with known parallaxes. Solar composition models by Chabrier \& Baraffe (1997) are also shown. Thin solid line refers to solar
metallicity VLM models transferred in the observational plane by using an
old set of colour - $T_{eff}$ transformations (see Cassisi et al. 2000).
Right panel: the fit of the MS stars population in a window of the bulge (Zoccali et al. 2000). The VLM models are the solar
metallicity ones adopted also in the left panel.}
\end{figure}

This result points out that the source of the shortcoming showed in fig.~3 (left panel) has not to be searched in the evolutionary
models but really in the adopted color - $T_{eff}$ relation: a drawback seems to exist in the evaluation 
of the opacity contribution at wavelenght $\lambda\le1\mu{m}$ in the computation of the model atmospheres.

This kind of results strongly suggest that evolutionary models for VLM stars have already attained a significant level of
accuracy in reproducing empirical constraints, and that a big improvement has to be expected in the adopted color -
$T_{eff}$ relations as a consequence of a more accurate treatment of the opacitive properties in model atmospheres
of cool and dense stellar objects.

\subsection{Diffusive processes}

Since many years, the constraints provided by Helioseismology have shown that atomic diffusion has
to be at work in the Sun. When relying on this circumstantial evidence, it is obvious to assume that this process has to be (more)
efficient also in metal-poor, low-mass stars (as a consequence of their thinner convective envelope).
However, this certainty is currently severerly challenged by spectroscopical measurements (Gratton et al. 2001, and references therein) 
showing that the iron abundance observed in stars belonging to metal-poor GCs are in disagreement with the
predictions provided by stellar models accounting for atomic diffusion: more in detail they found no significant differences between the
iron abundance observed in Turn-Off stars and RGB ones as one has to be expect in the case of efficient atomic diffusion.
However, recently Korn et al. (2006) have claimed the detection of a diffusion signature in the GC NGC~6397. According to their
analysis the TO stars disclose a lower [Fe/H] than the more evolved stars.

From the theoretical point of view, the problem of the efficiency of the various diffusive processes: atomic diffusion and radiative
levitation; has received a lot of attention thanks to the work by Richard et al. (2002), Vandenberg et al. (2002), and Richard et al.
(2005). Their sets of models, accounting simultaneously for atomic diffusion and levitation, predict that, at odds, with models accounting
only for atomic diffusion, the surface abundance of iron - and of the other heavy elements - is less depleted and it can be also become
overabundant - at the TO - due to radiative levitation (Richard et al. 2002).

However, one has to note that in order to achieve a fine agreement with both helioseismological constraints and
spectroscopical data for both field and clusters stars  as for instance for the Li trend with the $T_{eff}$ (the
so-called Spite plateau; Richard et al. 2005), some additional amount of mixing at the bottom of the canonical convective 
envelope has to be included.
It is common to refer to this extra-mixing as \lq{turbulence}\rq, but so far there is no firm physically grounded explanation for this
process.

Although the recent results provided by Korn et al. (2006) can be nicely interpreted in the framework of the diffusive ( $+$ turbulence) models of
Richard et al. (2005), there are at least two issues that should be considered: i) What is the physics behind the turbulence?, ii) Why for
the same GC, do independent groups find so different results? 
Is this occurrence due to a problem in the adopted $T_{eff}$ scale for metal-poor MS stars (see also Th{\'e}venin et al.~2001)?

A lot of theoretical work has to be done in order to understand all the physical mechanisms that can contribute to enhance or to decrease
the effects of diffusive processes. However it is also of pivotal importance to collect as many as possible independent spectroscopical
measurements in order to set on a more firm ground the observational scenario that has to be used in order to constraint the theoretical
framework.

\subsection{The Red Giant Branch evolutionary stage}

The RGB is one of the most prominent   
and well populated feature in the CMD of   
stellar populations with ages larger than about $1.5 - 2$ Gyr.    
The theoretical modeling of RGB stars plays therefore a wide ranging role,   
involving various fields of galactic and extragalactic astrophysics (see Salaris et al. 2002).
So, it is very important to verify the consistency between theoretical predictions
and observational data for RGB stars.

One of the major deficiencies in the stellar evolution theory is the lack of a a rigorous theory
of convection. As a consequence in the outer, super-adiabatic layers of stellar models, the mixing length theory 
is almost universally used. It contains a number of free parameters, whose numerical values affect the model $T_{\rm eff}$; one of them  
is $\alpha_{\rm MLT}$, the ratio of the mixing length to the pressure scale height, which provides the scale length of the convective motions:
for a given stellar luminosity it fixes the radius of the stellar model, and hence its $T_{eff}$. The $\alpha_{\rm MLT}$ parameter is commonly
calibrated by forcing stellar models of the Sun to reproduce the solar radius. However, the thermal conditions inside the envelope of RGB stars
are quite different in comparison with those existing in the solar envelope. So it is extremely important to check if RGB models whose  $\alpha_{\rm MLT}$
has been calibrated on the Sun, reproduce properly the empirical $T_{eff}$ of RGB stars. In this context the availability of
multi-bands photometry (from the
optical to the Near-Infrared ones), allowing an accurate determination of the effective temperature of cool stars as the RGB ones, is a
fundamental benchmark for stellar models.

Left Panel of fig.~4 shows the comparison between the most updated empirical database of $T_{eff}$ values for GC RGB stars from Ferraro et al. (2006) and
theoretical predictions as provided by some of the most recent stellar models libraries\footnote{We note that all these theoretical models 
are based on a solar calibration of the $\alpha_{\rm MLT}$.}. One can notice that almost all model predictions reproduce quite well the
empirical estimates in the whole metallicity range. However, at the same metallicity significant differences are present between the various
stellar models. This evidence has to be taken into account when comparing theory with observations in order to retrieve information about the
properties of a given stellar populations such as its metallicity. 

An other relevant issue for RGB stellar models is the uncertainty associated to the conductive opacity
adopted in model computations.
When the degree of electron degeneracy is significant, electron conduction is the 
dominant energy transport mechanism, and the value of the 
electron-conduction opacity enters the equation 
of the temperature gradient. This physical condition is verified in particular
in the He-core of low-mass stars during their RGB evolution (see Salaris et al. 2002).  The precise computation    
of the conductive opacities is fundamental for deriving the   
correct value of the He-core mass ($M_{cHe}$) at the He-flash, and hence the brightness of
both the RGB Tip and of the HB, i.e. two of the most important standard candles for Pop. II stellar systems. 

\begin{figure}
\centering
\resizebox{5.5cm}{!}{\includegraphics{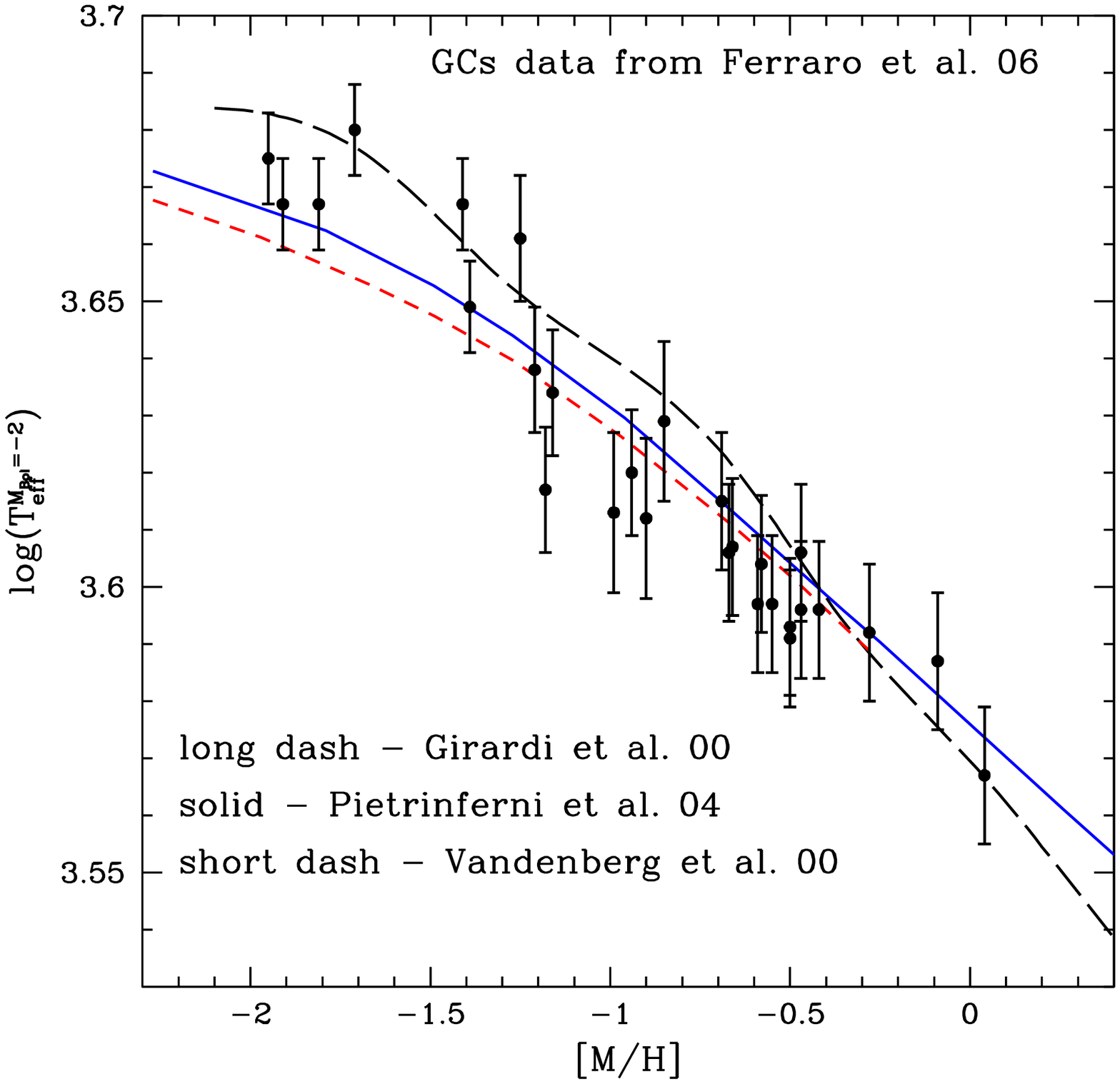} }
\resizebox{5.5cm}{!}{\includegraphics{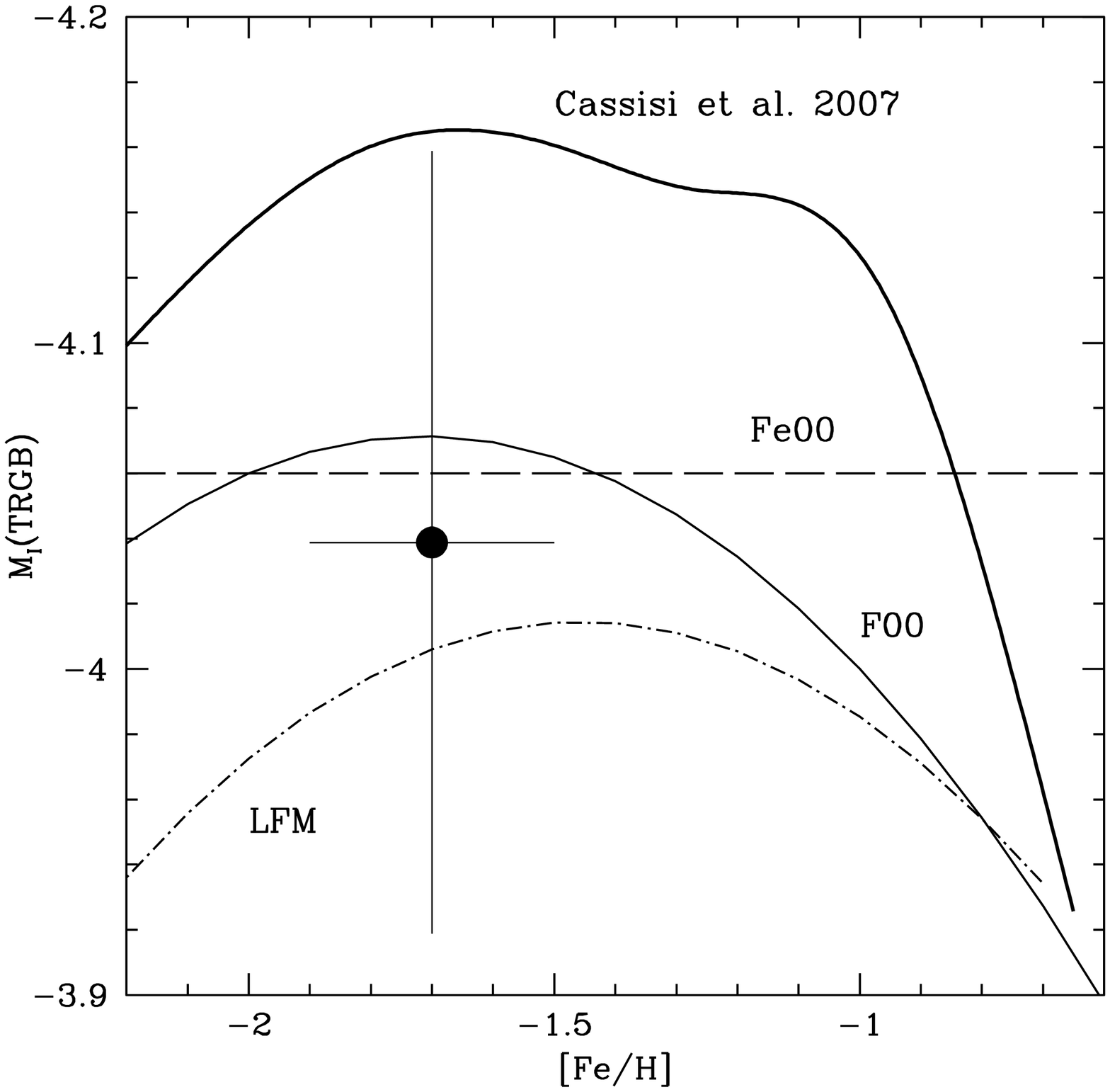} }
\caption[]{Left panel: RGB Effective temperature at $M_{Bol}=-2$ as a function of the metallicity for a sample of 
GGCs, with superimposed selected theoretical predictions from Girardi et al. (2000), Vandenberg et al. (2000) and Pietrinferni
et al. (2004).
Right panel: comparison between various calibrations of the TRGB
absolute I magnitude as a function of the
metallicity. The point with error bars corresponds to the estimate by Bellazzini et al. (2001) for $\omega$~Cen. 
The other calibrations are those by Ferrarese et al. (2000, ApJS, 128, 431 - Fe00), 
Ferraro et al. (2000, AJ, 119, 1282 - F00) and Lee et al. (1993, ApJ, 417, 553 - LFM). 
The theoretical calibration based on updated conductive opacity is shown as heavy solid line.}
\end{figure}

Regardless of the pivotal importance played by this \lq{ingredient}\rq, so far all available sources of conductive opacity
were affected by several limitations and shortcomings (see Catelan 2005).
Recently Cassisi et al. (2007) provided new predictions about the conductive opacity that largely improve in many aspects the old
results: the use of the new opacities cause a decrease of the value of $M_{cHe}$ of  $\sim0.006M_\odot$.

It is important to check if this improvement in the physics adopted for RGB model computations is supported by empirical constraints. For this
aim, figure~4 (right panel) shows a comparison between theoretical predictions about the I-Cousin band brightness of the RGB Tip as a function of the
metallicity and various empirical and semi-empirical calibrations. The new theoretical calibration of this standard candle appears to be in
agreement at the level of $1\sigma$ with the relevant empirical measurement in the GC $\omega$~Cen performed by Bellazzini et al. (2000).

\subsection{The Asymptotic Giant Branch}

The AGB stage is one of the most important evolutionary phases for several reasons: i) the nucleosynthesis, ii) AGB stars are reliable population
tracers, ii) its contribution to the integrated colors - mainly in the NIR bands - of unresolved stellar populations. 
From the theoretical point of view, the stellar models for AGB stars are really a challenging task. This is due to the evidence that the
evolutionary and structural properties of such objects are strongly affected by the strong link existing between the nuclear burning sources (the H- and
the He-burning shells), the mixing efficiency both in the envelope and in the inter-shell region, the opacitive properties of the envelopes
as well as the mass loss efficiency.

Concerning the mixing efficiency, the most important issue is related to the occurrence of the $3^o$ Dredge-up occurring in all stars with
initial mass larger than $1.2-1.4M_\odot$: we are not able to predict from first principles the efficiency of this mechanism and as a consequence
we are forced to use {\it ad hoc} parametrizations (Straniero et al. 2006). The main consequence of the $3^o$ Dredge-up is that a huge amount of Carbon is dredged
up the stellar surface and, indeed, $C/O$ values quite larger than unity - as really observed in C-stars - are achieved. When this occurs the
opacity properties of the star can not be longer described relying on opacity computed for scaled-solar heavy elements mixtures.

In this context, although many improvements have been obtained thanks to the work by Marigo (2002) and Marigo et al. (in preparation), we still lack of
robust predictions about the trend of radiative opacity for $C/O$ values of the order of unity or larger. This occurrence partially hampers our
capability to properly predict the effective temperature of AGB models and, hence, their color, as well as the mass-loss efficiency.

\begin{acknowledgments}
We warmly acknowledge G. Bono for an accurate reading of this manuscript. We are also grateful to 
G. Bono, S. Cristallo, M. Salaris, D. Vandenberg, A. Weiss for many enlightening discussions. We also wish to thank A. Pietrinferni 
for the help provided, as well as L. Girardi for providing results in advance of publication. 
\end{acknowledgments}

%
%
%

\begin{thebibliography}{}

\bibitem[]{} Allard, F., Hauschildt, P.H., Alexander, D.R. \& Starrfield, S. 1997, \textit{ARAA}, 35, 137

\bibitem[]{} Baraffe, I., Chabrier, G., Allard, F., \& Hauschildt, P.H., 1998, \textit{A\&A}, 337, 403

\bibitem[]{} Bellazzini, M., Ferraro, F.R., \& Pancino, E., 2001, \textit{ApJ}, 556, 635 

\bibitem[]{} Cassisi, S. 2004, in,  D.W. Kurtz and  K.R. Pollard (eds), \textit{Variable Stars in the Local Group}, 
Proc. IAU Colloquium 193, Vol. 310, p. 489

\bibitem[]{} Cassisi, S. 2005, in , D. Valls-Gabaud \& M. Ch\'avez (eds), Proc. of the meeting \textit{Resolved Stellar Populations}, 
in press (astro-ph/0506161) 

\bibitem[]{} Cassisi, S., Castellani, V., Ciarcelluti, P., Piotto, G. \& Zoccali, M. 2000, \textit{MNRAS}, 315, 679

\bibitem[]{} Cassisi, S., Castellani, V., Degl'Innocenti, S., Salaris, M., \& Weiss, A. 1999, \textit{A\&AS}, 134, 103

\bibitem[]{} Cassisi, S., Castellani, V., Degl'Innocenti, S., \& Weiss, A. 1998, \textit{A\&AS}, 129, 267

\bibitem[]{} Cassisi, S., Potekhin, A., Pietrinferni, A., Catelan, M. \& Salaris, S. 2007, \textit{ApJ}, submitted to

\bibitem[]{} Castellani, V. 1999, in, Martines Roger, C. et al. (eds), \textit{Globular clusters}, p.109

\bibitem[]{} Castellani, V., \& Degl'Innocenti, S. 1999, \textit{A\&A}, 344, 97

\bibitem[]{} Catelan, M. 2005, in, D. Valls-Gabaud \& M. Ch\'avez (eds), Proc. of the meeting \textit{Resolved Stellar Populations}, 
in press (astro-ph/0507464) 

\bibitem[]{} Chaboyer, B. 1995, \textit{ApJ}, 444, L9

\bibitem[]{} Chabrier, G. \& Baraffe, I. 2000, \textit{ARAA}, 38 337

\bibitem[]{} Cox, J.P., \& Giuli, R.T. 1968, \lq{Principles of stellar structure}\rq,  New York, Gordon and Breach 

\bibitem[]{} Ferraro, F.R., Valenti, E., Straniero, O. \& Origlia, L. 2006,\textit{ApJ}, 642, 225

\bibitem[]{} Gallart, C., Zoccali, M. \& Aparicio, A. 2005, \textit{ARAA}, 43, 387

\bibitem[]{} Girardi, L., Bressan, A., Bertelli, G. \& Chiosi, C. 2000, \textit{A\&AS}, 141, 371

\bibitem[]{} Gratton, R.G., et al. 2001, \textit{A\&AS}, 369, 87

\bibitem[]{} King, I.R., Anderson, J., Cool, A.M. \&  Piotto, G. 1998, \textit{ApJ}, 492, L37

\bibitem[]{} Kippenhahn, R., \& Weigert, A. 1990, \lq{Stellar structure and evolution}\rq, Springer-Verlag

\bibitem[]{} Korn, A.J., et al. 2006, \textit{Nature}, 442, 657

\bibitem[]{} Krishna-Swamy, K.S., 1966, \textit{ApJ}, 145, 174   

\bibitem[]{} Marigo, P. 2002, \textit{A\&A}, 387, 507

\bibitem[]{} Pietrinferni, A., Cassisi, S., Salaris, M. \& Castelli, F. 2004, \textit{ApJ}, 612, 168 

\bibitem[]{} Richard, O., Michaud, G., Richer, J., Turcotte, S., Turck-Chi\'eze, S., \& VandenBerg, D.A. 2002, \textit{ApJ}, 568, 979

\bibitem[]{} Richard, O., Michaud, G. \& Richer, J. 2005, \textit{ApJ}, 619, 538

\bibitem[]{} Salaris, M. \& Cassisi, S. 2005, \lq{Evolution of stars and stellar populations}\rq, Wiley \& sons

\bibitem[]{} Salaris, M., Cassisi, S., \& Weiss, A. 2002, \textit{PASP}, 114, 375

\bibitem[]{} S\'egransan, D., Kervella, P., Forveille, T. \& Queloz, D. 2003, \textit{A\&A} 397, L5

\bibitem[]{} Straniero, O., Gallino, R. \& Cristallo, S. 2006, \textit{Nucl. Phys. A}, 777, 311

\bibitem[]{} Th{\'e}venin, F. et al. 2001, \textit{A\&A}, 373, 905

\bibitem[]{} Vandenberg, D.A., Hartwick, F.D.A., Dawson, P. \& Alexander, D.R., 1983, \textit{ApJ}, 266, 747

\bibitem[]{} VandenBerg, D.A., Richard, O., Michaud, G., \& Richer, J. 2002, \textit{ApJ}, 571, 487

\bibitem[]{} VandenBerg, D.A., Swenson, F.J. Rogers, F.J., Iglesias, C.A., \& Alexander, D.R. 2000, \textit{ApJ}, 532, 430

\bibitem[]{} Zoccali, M., et al. 2000, \textit{ApJ}, 530, 418

\end{thebibliography}
\end{document}